\documentclass[aps,twocolumn,groupedaddress,floatfix,longbibliography,superscriptaddress]{revtex4-1}
\usepackage{amsmath}
\usepackage{amssymb}
\usepackage{amsfonts}
\usepackage{graphicx}
\usepackage{bbold}
\usepackage{bm}
\usepackage{bm}
\usepackage{times,float}
\usepackage{graphicx}
\usepackage[usenames,dvipsnames,svgnames]{xcolor}
\usepackage{hyperref}
\hypersetup{colorlinks=true, linkcolor=Blue, citecolor=Green,urlcolor=Blue}

\newcommand{\be}{\begin{equation}}
\newcommand{\ee}{\end{equation}}
\newcommand{\bea}{\begin{eqnarray}}
\newcommand{\eea}{\end{eqnarray}}
\newcommand{\beas}{\begin{eqnarray*}}
\newcommand{\eeas}{\end{eqnarray*}}
\newcommand{\nn}{\nonumber}

\newcommand{\eq}[1]{Eq.~(\ref{#1})}

\newcommand{\hrho}{\hat{\rho}}
\newcommand{\hvrr}{\hat{\varrho}}

\begin{document}
\title{{Super-statistics and quantum entanglement in the isotropic spin-1/2 XX dimmer from a non-additive thermodynamics perspective}}
\author{Jorge David Casta\~no-Yepes}\email[Corresponding author: ]{ jorgecastanoy@gmail.com}
\affiliation{Facultad de Ciencias - CUICBAS, Universidad de Colima, Bernal D\'iaz del Castillo No. 340, Col. Villas San Sebasti\'an, 28045 Colima, Mexico.}
\author{Cristian Felipe Ramirez-Gutierrez}
\affiliation{Universidad Polit\'ecnica de Quer\'etaro, El Marqu\'es,   76240 Quer\'etaro, Mexico.}
\begin{abstract}
In this paper, the impact of temperature fluctuations in the entanglement of two qubits described by { a spin-1/2 XX model} is studied. To describe the out-of-equilibrium situation, super-statistics is used with fluctuations given by a $\chi^2$ distribution function, and its free parameters are chosen in such a way that resembles the non-additive Tsallis thermodynamics. In order to preserve the Legendre structure of the thermal functions, particular energy constraints are imposed on the density operator and the internal energy. Analytical results are obtained using an additional set of constraints after a parametrization of the physical temperature. We show that the well-known parametrization may lead to undesirable values of the physical temperature so that by analyzing the entropy as a function of energy, the correct values are found. The quantum entanglement is obtained from the Concurrence and is compared with { the case when the Tsallis restrictions are not imposed on the density operator.}
\end{abstract}
\maketitle
\onecolumngrid
\section{Introduction}\label{Sec:Intro}

Super-statistics (SS), which were developed by Beck and Cohen, is an effective way to study the non-equilibrium processes based on the idea that the thermodynamics of a system that passes through several equilibrium thermodynamic states or is divided into subsystems, each in local equilibrium, can be approximated by postulating an average Boltzmann factor weighted with the density probability function of fluctuations~\cite{beck2003superstatistics,beck2004superstatistics,beck2009recent}. The SS framework has applied to many topics: turbulence models~\cite{PhysRevLett.91.084503,PhysRevE.72.026304}, cosmic rays~\cite{beck2004generalized}, quantum cromodynamics~\cite{PhysRevD.98.114002,PhysRevD.91.114027}, quantum dots~\cite{castano2020super,sargolzaeipor2019superstatistics}, out-of-equilibrium Ising models~\cite{PhysRevE.103.032104}, quantum field theory~\cite{ishihara2018momentum,ishihara2018phase}, among others~\cite{PhysRevC.79.054903,PhysRevE.88.062146,PhysRevD.95.124031,bediaga2000nonextensive}. Then, SS is a subject that deserves attention.

The present work has a particular interest in the paper by K. Ourabah and M. Tribeche, where the authors explore the impact of temperature fluctuations in the entanglement of quantum systems provided by the concurrence function~\cite{PhysRevE.95.042111}. In order to model such fluctuations, the SS framework is implemented by introducing different distribution functions. The authors found effects of the temperature field variations into the thermal superstates, where non-equilibrium assumptions determine if the entanglement is enhanced or prevented. 
Thus, we present an extension of those insights, based on the so-called $\chi^2$-distribution function is closely related to the non-additive Tsallis formalism (for example, see Ref.~\cite{PhysRevE.101.040102}), and we explore the consequences of its restriction over quantum phenomena.\\

The formal analysis of SS is based on fluctuations of an intensive parameter $\tilde{\beta}$ (inverse temperature, chemical potential, noise, mass variation, volatility in finance, etc), which are described by a probability density function $f(\tilde{\beta})$. If the system admits states of equilibrium described by Boltzmann-like probabilities, the SS description is addressed with an average over such stages, given by
\bea
\hat{B}=\int_0^\infty d\tilde{\beta} f(\tilde{\beta}) e^{-\tilde{\beta}\hat{H}},
\label{Bdef},
\eea
where $\hat{H}$ is the Hamiltonian of the system and $\hat{B}$ is known as the averaged or modified Boltzmann factor. For analytical results, it is common to identify $f(\tilde{\beta})$ with the $\chi^2$ or Gamma distribution function, namely
\bea
f(\tilde{\beta})=\frac{1}{b\Gamma(c)}\left(\frac{\tilde{\beta}}{c}\right)^{c-1}e^{-\tilde{\beta}/b},
\label{chisquared}
\eea
where $b$ and $c$ are free parameters. Tsallis-like description is found when the parameters are restricted to $bc=\beta$ and $c=1/(q-1)$, so that the modified Boltzmann factor is
\bea
\hat{B}=e_q^{-\beta\hat{H}}
\label{B1eq}
\eea
where
\bea
e_q^x\equiv[1+(1-q)x]^{1/(1-q)},
\eea
which after a proper normalization, resembles the non-additive probability weight factor~\cite{tsallis1998role}. { In the last equations, the parameter $\beta$ is identified with the average temperature through the expression~\cite{beck2003superstatistics,beck2004superstatistics,beck2009recent}:
\bea
\int_0^\infty d\Tilde{\beta} f(\Tilde{\beta})\Tilde{\beta}=\beta.
\eea}

As it was shown in previous works~\cite{castano2020super,castano2020comments}, admitting parallelism between SS and non-extensive thermodynamics has profound implications in the energy constraints to let invariant the Legendre structure (LS) behind a thermal treatment. 

Although the computation of quantum entanglement in Ref.~\cite{PhysRevE.95.042111} does not use the LS (in the sense of computing observables from derivatives of the natural logarithm of partition function), it is well established that the validity of LS related to \eq{B1eq} implies a correct identification of the physical temperature, which is different from the authors one.

In this work, the impact of the temperature redefinition on quantum entanglement is achieved by following the results of the Tsallis non-additive theory. We point out that the presented results are not a correction of Ourabah and Tribeche's paper: they extend it. The last is because, in that research, the implementation of the $\chi^2$-distribution function is never associated with a non-additive formalism, which in general, is correct. Hence, {{\it the present discussion is valid when SS with $\chi^2$-distribution function with the particular parameters choice for $c$ and $b$ is identified with the Tsallis statistics.}}

The paper is organized as follows: Sec.~\ref{Sec:ReviewTsallis} presents a short review of the Tsallis non-extensive formalism and the physical temperature definition. In Sec.~\ref{Sec:XYmodel} { the model for a spin-1/2 XX dimmer} is introduced and the thermal state is found. In Sec.~\ref{sec:physicaltemp} we show that the parametrization of the physical temperature leads to undesirable values of $T$ and by analyzing the entropy as a function of internal energy, we found restrictions over that parametrization. Section~\ref{Sec:Concurrence} is devoted to the Concurrence, as a measure of quantum entanglement. In that section, the concurrence for a non-additive point of view is compared with the Gibbs-Boltzmann and with the framework where SS is not identified with Tsallis formalism. Finally, the summary and conclusions are presented in Sec.~\ref{Sec:summary}.

\section{Thermal state in the Tsallis formalism}\label{Sec:ReviewTsallis}
In order to generalize the Gibbs-Boltzmann (GB) statistics, Tsallis proposed an entropy function depending on a parameter $q$~\cite{tsallis1988possible}. As a postulate, it takes the form ($k_\text{B}=1$):
\bea
S=\frac{1}{q-1}\left(1-\text{Tr}\left[\hrho^q\right]\right)\forall\;q \in\mathbb{R},
\label{TsallisEntropy}
\eea
were $\hat{\rho}$ is the thermal state or density operator. To connect with the results of the usual Boltzmann statistics, Tsallis argued that the preservation of the LS leads to a particular energy constrain given by~\cite{tsallis1998role}:
\bea
U=\frac{\text{Tr}\left[\hrho^q\hat{H}\right]}{\text{Tr}\left[\hrho^q\right]},
\label{U}
\eea
where $\hat{H}$ is the Hamiltonian of the system. Then, after maximizing $S$ is is straightforward to find that
\bea
\hrho=\frac{1}{Z}\exp_q\left(-\beta\frac{\hat{H}-U}{\text{Tr}\left[\hrho^q\right]}\right),
\label{rho}
\eea
where $\beta$ is the inverse physical temperature and the partition function $Z$ is
\bea
Z=\text{Tr}\left[\exp_q\left(-\beta\frac{\hat{H}-U}{\text{Tr}\left[\hrho^q\right]}\right)\right].
\label{Z}
\eea

It is worth saying that LS can be demanded by more fundamental arguments related to an increasing entropy and a positive definite specific heat~\cite{scarfone2016consistency,plastino1997universality,castano2020super}. Moreover, the thermodynamic functions can be expressed in terms of $Z$ as follows
\begin{subequations}
\bea
{ F(\beta)}\equiv U(\beta)-TS=-\beta^{-1}\ln_qZ_1,
\label{Omega}
\eea
\bea
U=-\partial_\beta \ln_qZ_1,
\label{U2}
\eea
where $Z_1$ is defined from
\bea
\ln_q Z_1=\ln_q Z-\beta U.
\eea
\end{subequations}

It is clear that Eqs.~(\ref{U})-(\ref{Z}) { for $\hrho$ are given in an implicit way}, which in general, have not a trivial solution. Nevertheless, they can be related to a second energy constraint proposal that preserves the LS which is given in terms of a non-physical temperature $\beta^\star$. Explicitly, {the thermal state $\hvrr$ and internal energy $\mathcal{U}$ for the second constraints are}
\bea
\hvrr=\frac{1}{\mathcal{Z}}\exp_q\left(-\beta^\star\hat{H}\right),
\label{rho2}
\eea
and
\bea
\mathcal{U}=\text{Tr}\left[\hvrr^q\hat{H}\right]=-\partial_{\beta^\star}\ln_q\mathcal{Z},
\eea
where
\bea
\mathcal{Z}=\text{Tr}\left[\exp_q\left(-\beta^\star\hat{H}\right)\right].
\label{Z2}
\eea

{ Taking into account the last equations, the relation between Eqs.~(\ref{U})-(\ref{Z}) and Eqs.~(\ref{rho2})-(\ref{Z2}) are~\cite{tsallis1998role}:}
\bea
\hrho(\beta)&=&\hvrr(\beta^\star)\nn\\
Z_1(\beta)&=&\mathcal{Z}(\beta^\star), 
\label{Z1andZ}
\eea
so that the physical temperature is parametrized with $\beta^\star$ as follows 
\bea
\beta=\frac{\beta^\star\,\text{Tr}\left[\hvrr^q(\beta^\star)\right]}{1-(1-q) \beta^\star \mathcal{U}\left(\beta^\star\right) / \text{Tr}\left[\hvrr^q(\beta^\star)\right]}.
\label{Eq:RenormalizedBeta2}
\eea

Therefore, it is possible to find thermal-related observables from $\hvrr$ and $\mathcal{Z}$ as a function of $\beta^\star$. For example, the physical entropy and internal energy will be
\begin{subequations}
\bea
S=-\frac{\left(\beta^\star\right)^2}{\partial_{\beta^\star}\beta}\partial_{\beta^\star}\left[\beta^{-1}\ln_q\mathcal{Z}(\beta^\star)\right]
\eea
\bea
U=-\left(\partial_{\beta^\star}\beta\right)^{-1}\partial_{\beta^\star}\ln_q\mathcal{Z}(\beta^\star),
\eea
\label{SandU}
\end{subequations}
so that after applying Eq.~(\ref{Eq:RenormalizedBeta2}), { the physically meaningful $S$ and $U$ are obtained.}

{
\section{The spin-1/2 XX dimmer}\label{Sec:XYmodel}
}
{ A model for a spin-1/2 XX dimmer or the quantum Heisenberg anti-ferromagnetic XY model for two interacting qubits} in the presence of an external magnetic field $B$ is described by the Hamiltonian~\cite{PhysRevA.64.012313}
\bea
\hat{H}=-J\left(\sigma_1^+\sigma_2^-+\sigma_1^-\sigma_2^+\right)+\frac{B}{2}\left(\sigma_1^z\otimes\sigma_0+\sigma_0\otimes\sigma_2^z\right),\nn\\
\label{Hamiltonian1}
\eea
where $J$ is the exchange interaction between the two spins while $J>0$ and $J<0$ correspond to the antiferromagnetic and ferromagnetic cases, respectively, and
\bea
\sigma_1^\pm\equiv\sigma^+\otimes\sigma_0,\hspace{0.5cm}\sigma_2^\pm\equiv\sigma_0\otimes\sigma^\pm,\hspace{0.5cm}\sigma_0=\mathbb{1}_{2\times2}.\nn\\
\eea


The igenvalues and eigenvectors of the Hamiltonian are
\bea
\hat{H}|00\rangle&=&-B|00\rangle\nn\\
\hat{H}|11\rangle&=&B|11\rangle\nn\\
\hat{H}\left(\frac{|01\rangle\pm|10\rangle}{\sqrt{2}}\right)&=&\pm J \left(\frac{|01\rangle\pm|10\rangle}{\sqrt{2}}\right),
\eea
then, in the basis $\left\{|00\rangle,|11\rangle,|01\rangle\right\},|10\rangle$ it is straightforward to find that
{
\bea
\!\!\!\hat{\varrho}=\frac{1}{Z}\begin{pmatrix}
e_q^{-\beta^\star B} & 0 & 0 & 0\\ 
0 & \cosh_q(\beta^\star J) & -\sinh_q(\beta^\star J) & 0 \\ 
0 & -\sinh_q(\beta^\star J) & \cosh_q(\beta^\star J) & 0\\ 
0 & 0 & 0 & e_q^{\beta^\star B}
\end{pmatrix},
\label{varrho}
\eea}
where
\bea
\cosh_q(x)&=&\frac{e_q^x+e_q^{-x}}{2},\nn\\
\sinh_q(x)&=&\frac{e_q^x-e_q^{-x}}{2},
\eea
and
{
\bea
Z=2\left[\cosh_q(\beta^\star J)+\cosh_q(\beta^\star B)\right].
\eea}

\section{The physical temperature}\label{sec:physicaltemp}
One challenging aspect in the Tsallis and $\chi^2$ formalisms is to give an interpretation for the $q$-parameter. By admitting that SS is an average over $M$ configurations of the intensive quantity. It can be argued that $q=1+2/M$ and $q>1$ so that in the ``thermodynamic" limit ($M\to\infty$) $q=1$, and the Gibbs-Boltzmann statistics is recovered~\cite{castano2020super,PhysRevD.98.114002}. Nevertheless, there is no {\it a priori} reason to ignore values of $q<1$. Here we explore both avenues.

\begin{figure*}[h]
    \centering
    \includegraphics[scale=0.455]{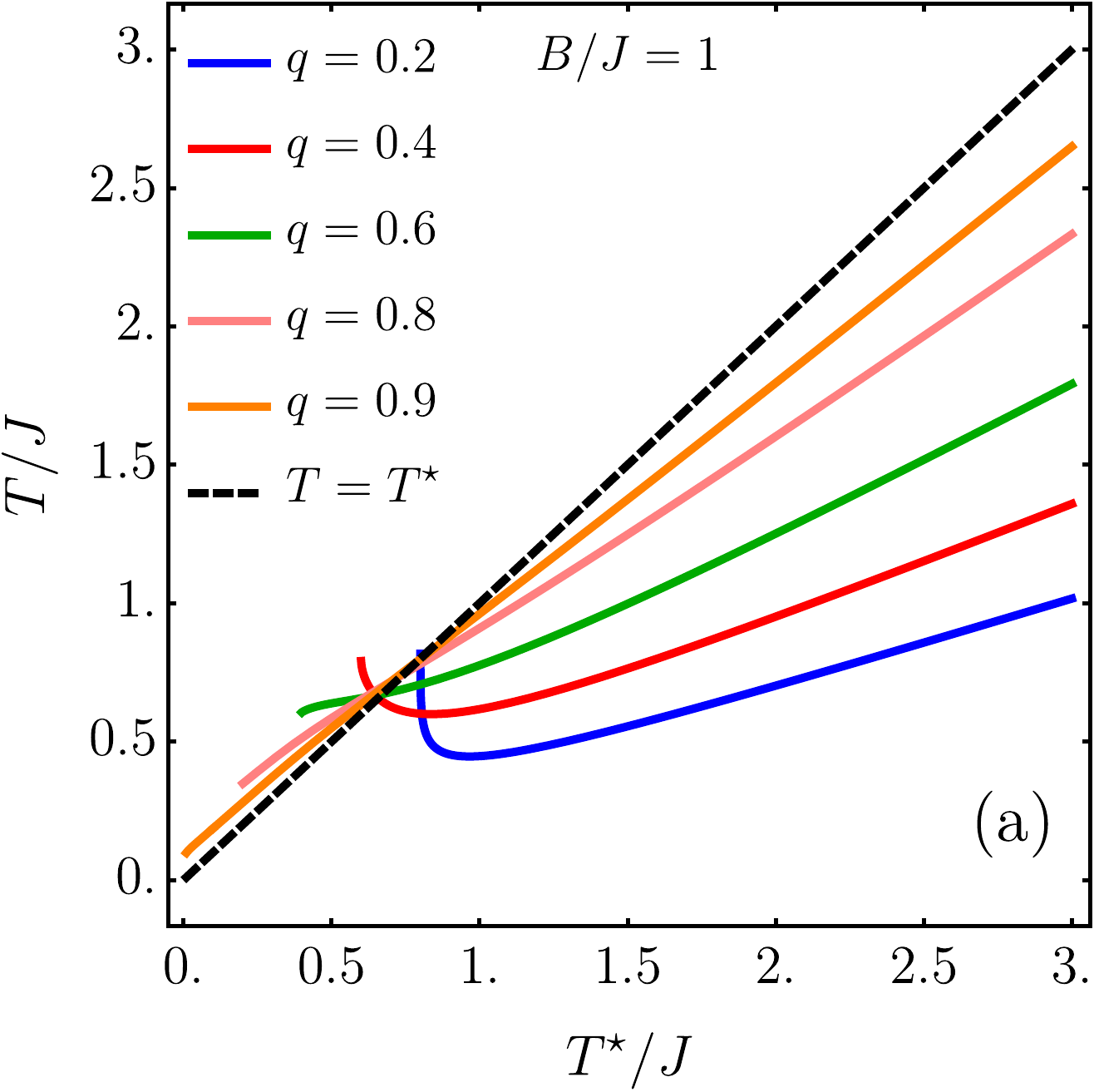}\hspace{0.25cm}
    \includegraphics[scale=0.455]{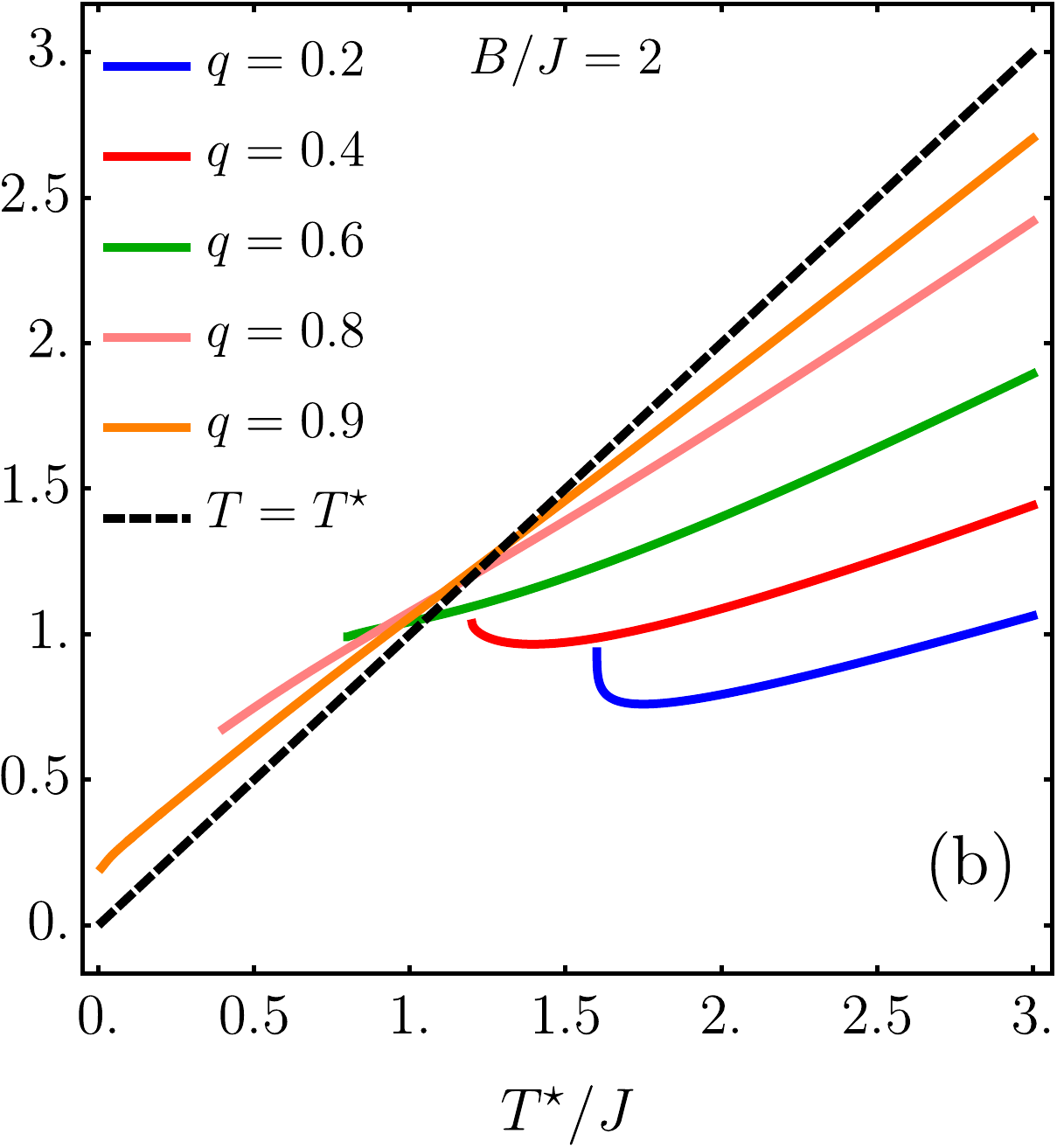}\hspace{0.25cm}
    \includegraphics[scale=0.455]{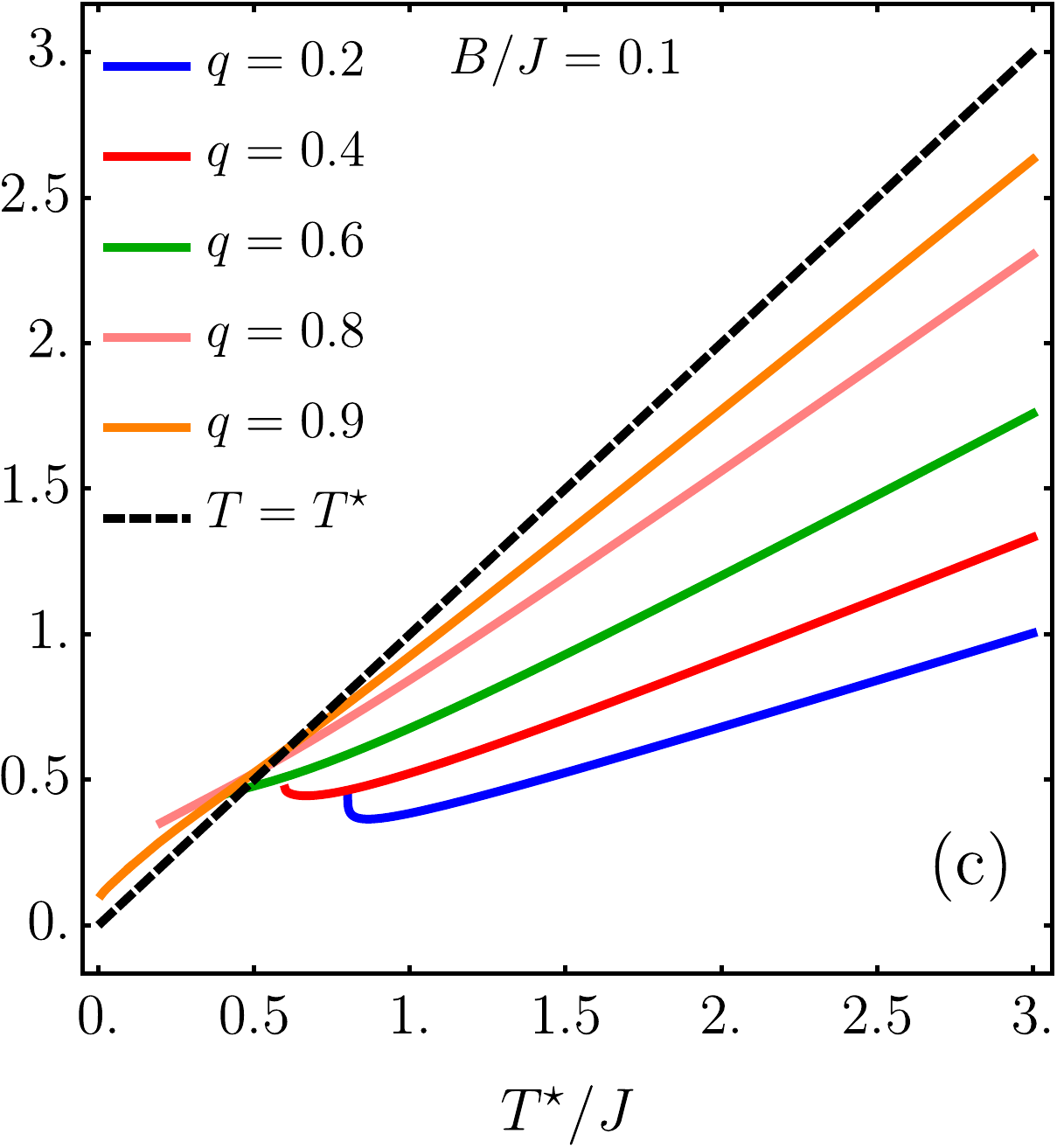}
    \caption{(Color online) Physical temperature $T=1/\beta$ as a function of the pseudo-temperature $T^\star=1/\beta^\star$ (both in units of $J$) computed from \eq{Eq:RenormalizedBeta2} for several values of $q<1$ and $B/J$. The Gibss-Boltzmann limit $T=T^\star$ or $q\to1$ (dashed line) is plotted for comparison purposes.}
    \label{Fig:TvsT1}
\end{figure*}

The physical temperature $T=1/\beta$ computed from \eq{Eq:RenormalizedBeta2} is shown in Fig.~\ref{Fig:TvsT1}, as a function of the pseudo-temperature $T^\star=1/\beta^\star$ provided by the thermal state of \eq{varrho} with $q<1$. As can be noticed, depending on the values of the parameters, not all the temperatures are reached by the system. In principle, it has a relation with the argument of $q$-functions were $1+(1-q)x\geq0$. This effect is know in the so-called kappa-distributions, related to correlated systems out of equilibrium, which are close to the non-additive formalism~\cite{livadiotis2015introduction,livadiotis2009beyond}. The latter has implications in the temperature notion and the connection to statistical mechanics: correlations between particles make temperature bounded below~\cite{livadiotis2013evidence}. Then, the curves suggest that the system admits the definition of temperature until a specific minimum value for a particular parameters window. Of course, restrictions over $\beta^\star$ or $J$ and $B$ are present in Ref.~\cite{PhysRevE.95.042111}, but as Fig.~\ref{Fig:TvsT1} shows, the actual temperature value differs from what they had considered.

\begin{figure}[h]
    \centering
    \includegraphics[scale=0.55]{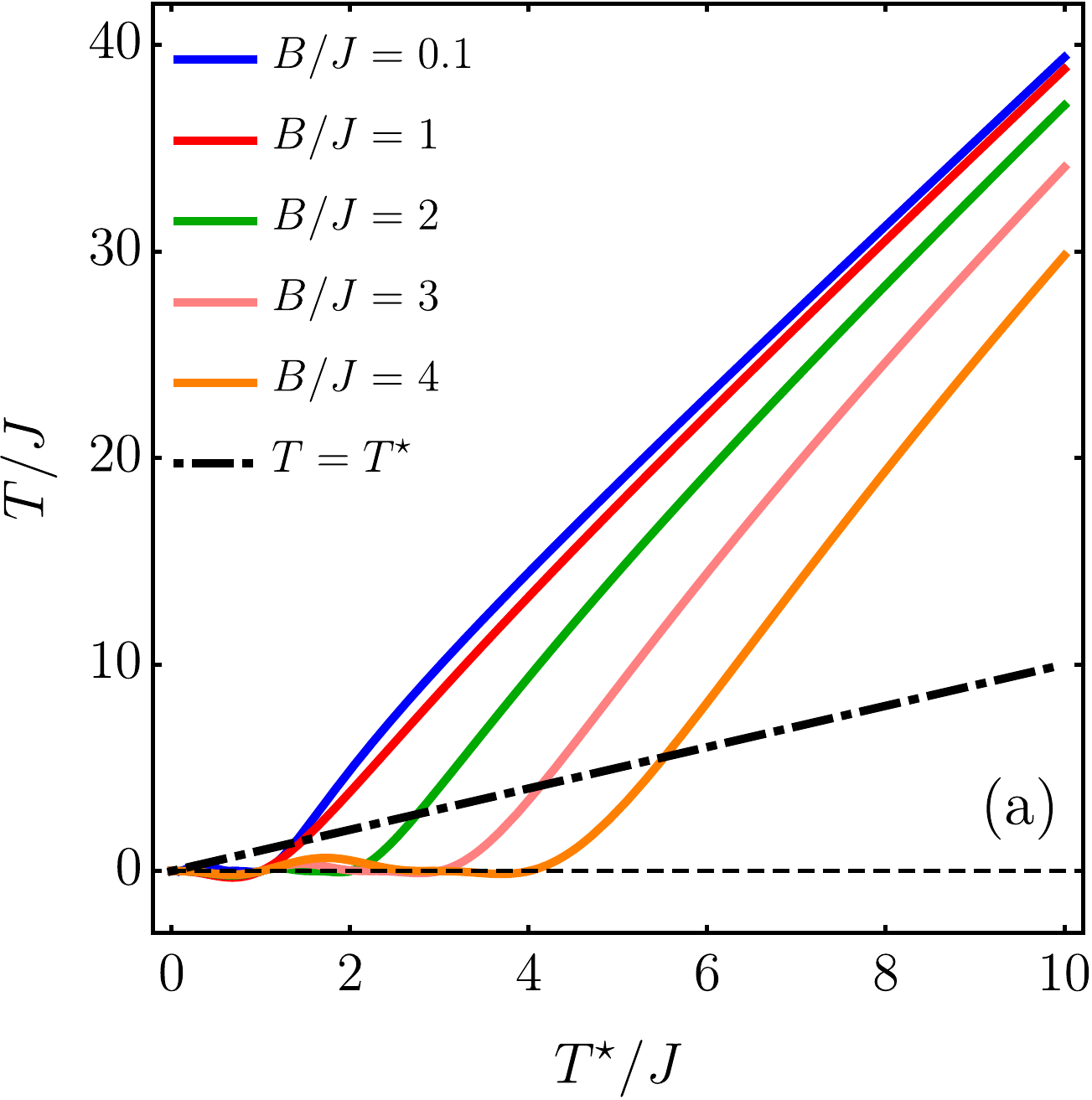}\hspace{1cm}\includegraphics[scale=0.55]{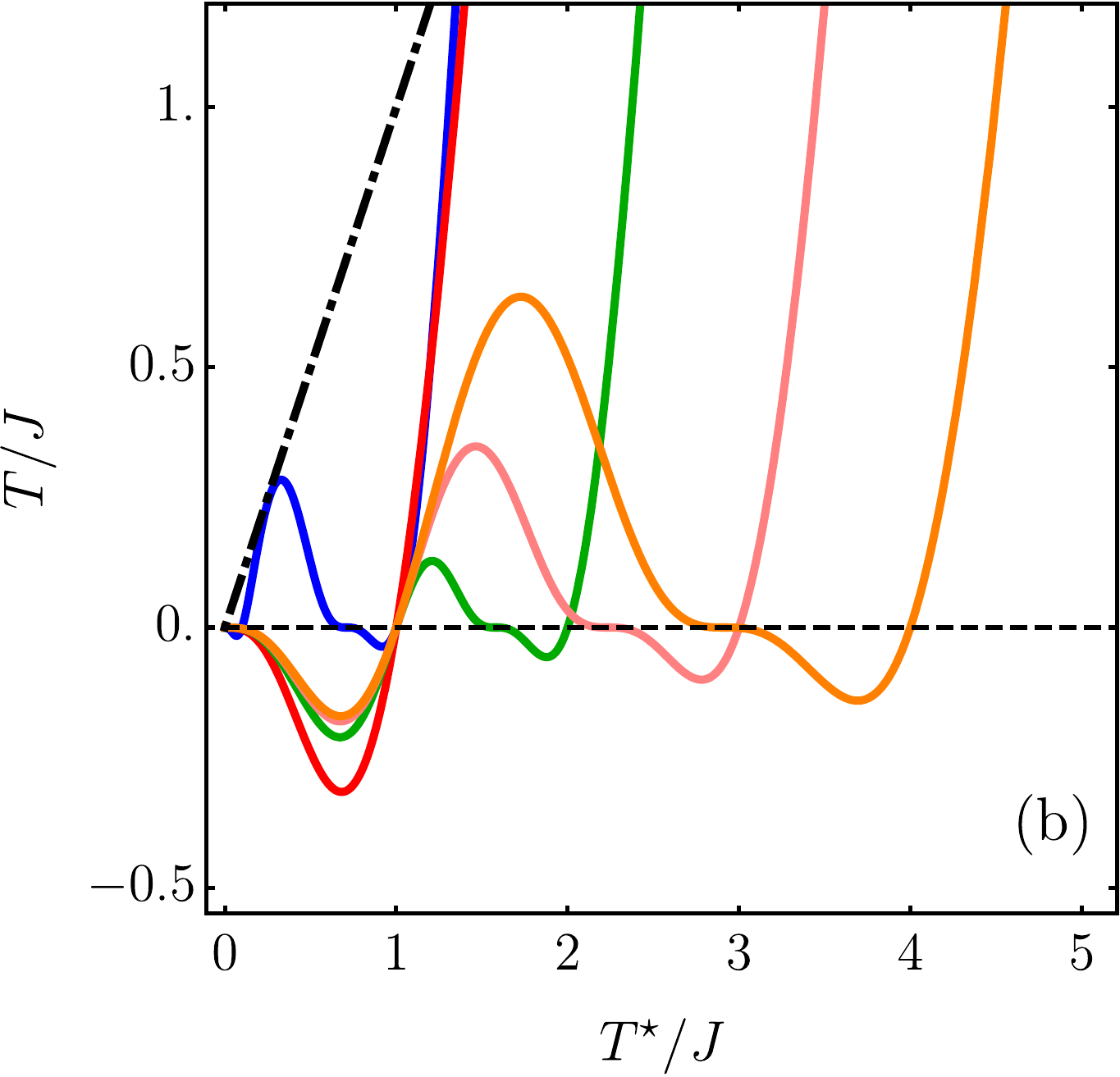}
    \caption{(Color online) (a) Physical temperature $T=1/\beta$ as a function of the pseudo-temperature $T^\star=1/\beta^\star$ (both in units of $J$) computed from \eq{Eq:RenormalizedBeta2} for $q=2$ and several values of $B/J$. The Gibss-Boltzmann limit $T=T^\star$ or $q\to1$ (dashed line) is plotted for comparison purposes. (b) Magnification in order to appreciate the prediction of negative physical temperatures.}
    \label{Fig:TvsT1q2}
\end{figure}

In contrast, for the case $q>1$ consider the particular value $q=2$: the differences are more significant and provide another physical point of view. Figure~\ref{Fig:TvsT1q2} shows the physical temperature when the quotient $B/J$ changes, with the particular feature that depending on the interactions strength, the system can reach negative absolute temperatures. That is not a surprising result, given that eigenvalues of the Hamiltonian are bounded from above. In the limit of high $B$ or low $J$, the system behaves { as the limit of the Ising model in which the interaction term becomes negligible. Therefore, it is close to a two-level system which develops absolute negative temperatures}~\cite{pathria}. Nevertheless, it marks a crucial difference between the formalisms, and it can be uses to test the validity of the non-additive perspective of temperature fluctuations.

Another critical point to discuss is that the parametrization of Eq.~(\ref{Eq:RenormalizedBeta2}) gives wrong values of $T$, even with the restriction $1+(1-q)x\geq0$ imposed to $\mathcal{Z}$. To clarify this point, note that the parametrization does not return a physical temperature $T$ monotonically increasing for all the values of $T^\star$, which implies that $\beta^\star$ is not a single valued function of $\beta$. Then, when the physical thermal observables are extracted from the relations of Eq.~(\ref{Z1andZ}), they are not a function of $T$, in the sense of the vertical line test.

\begin{figure*}[h]
    \centering
    \includegraphics[scale=0.52]{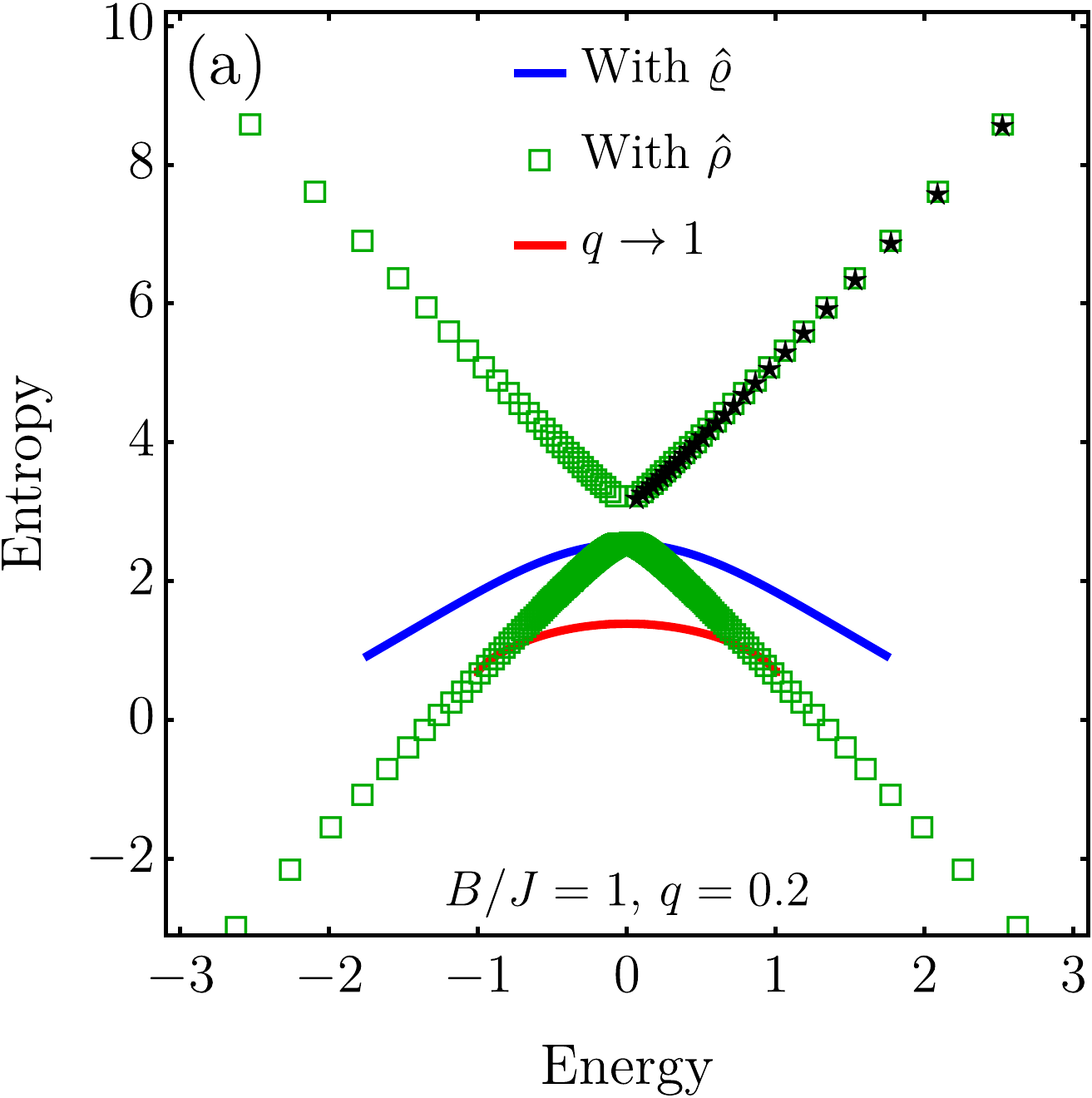}
    \hspace{0.4cm}
    \includegraphics[scale=0.52]{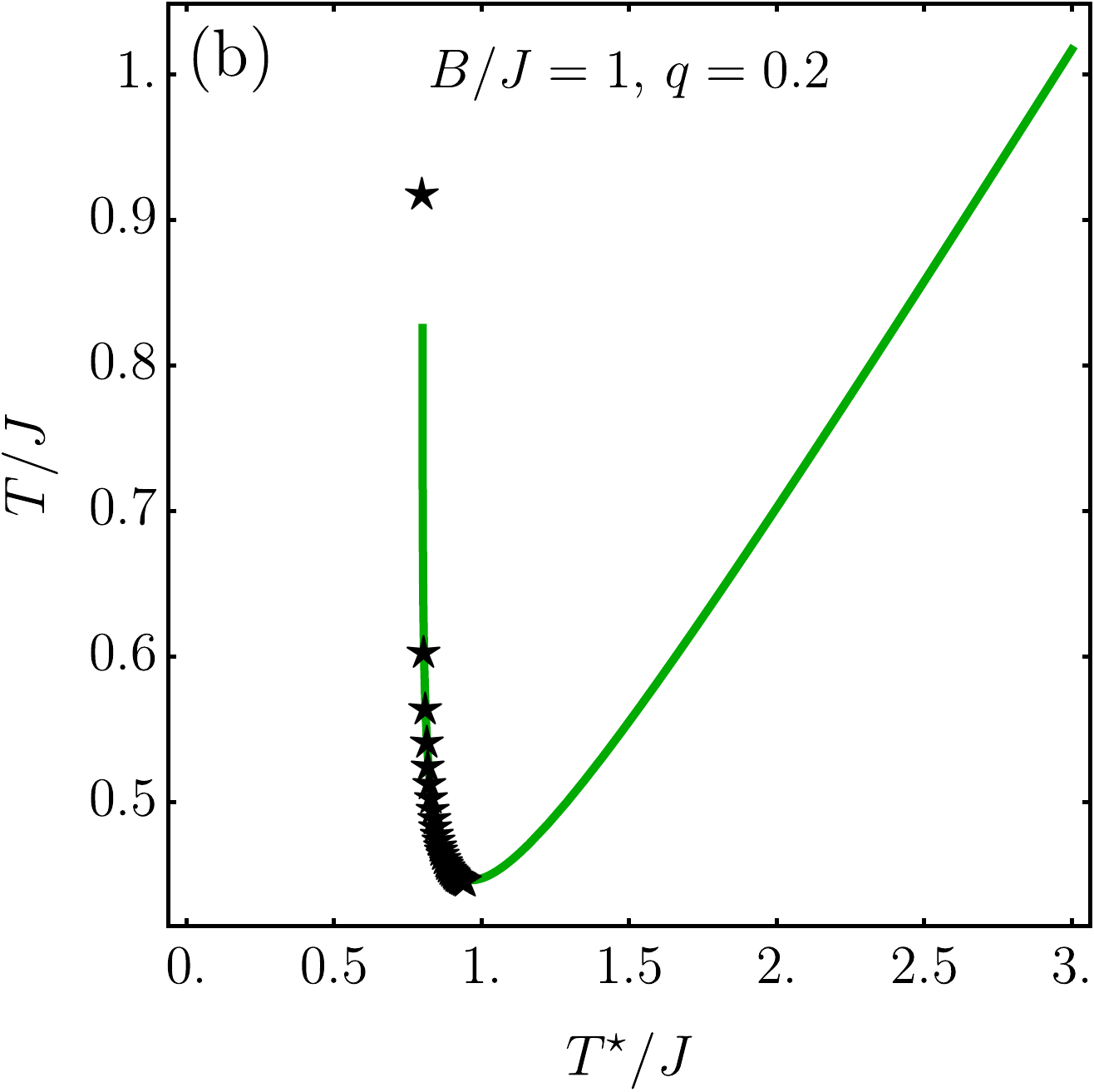}\\
    \vspace{0.5cm}
    \includegraphics[scale=0.52]{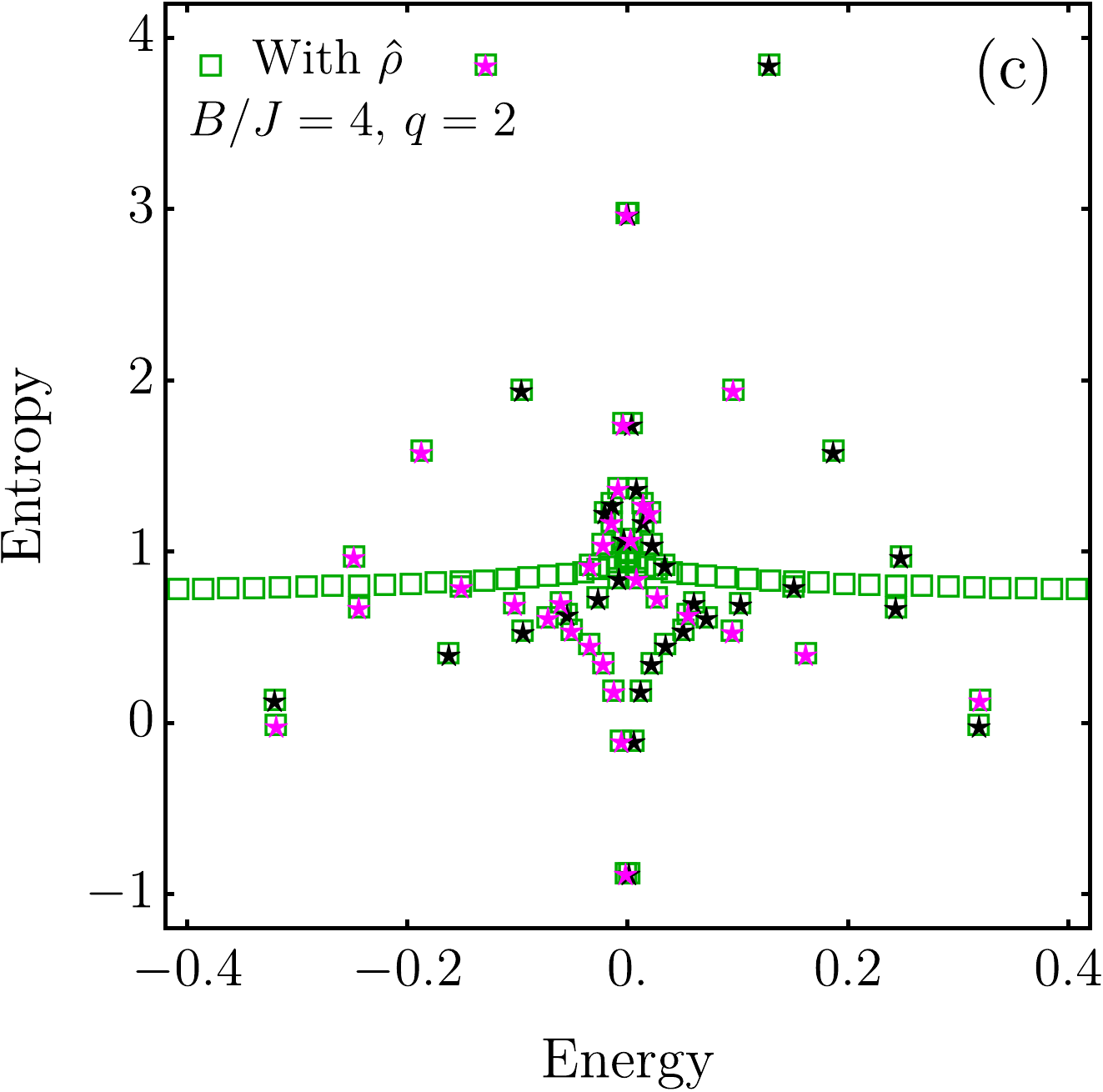}\hspace{0.4cm}
    \includegraphics[scale=0.52]{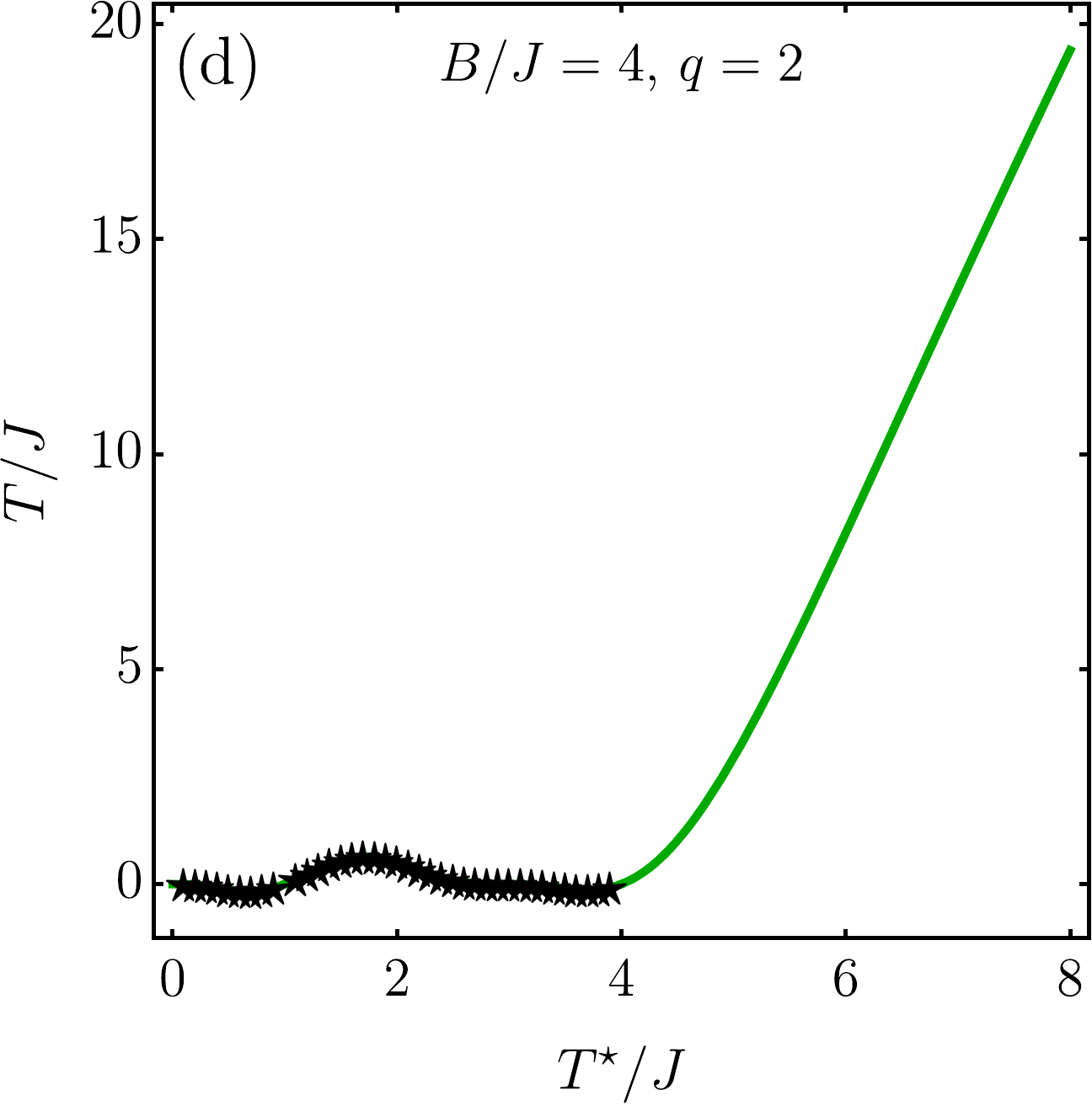}
    \caption{(Color online) (a) Entropy as a function of the internal energy, for the non-extensive formalism of Eqs.~(\ref{TsallisEntropy}) and ~(\ref{U}) (green squares), compared with the formalism related to $\hvrr$ (blue line) and the GB case $q\to1$ (red line) for $B/J=1$ and $q=0.2$. The black stars are non-physical (positive) temperatures provided by parametrization of Eq.~(\ref{Eq:RenormalizedBeta2}), where the thermal observables are not functions of $T$. (b) Identification of the non-physical values of $T$ by using panel (a). In panel (c) $S$ vs. $U$ is computed from $\hrho$ for $B/J=4$ and $q=2$. The black (pink) stars are the non-physical positive (negative) parametrized temperatures which are identified in panel (d).}
    \label{Fig:SandUab}
\end{figure*}

In order to know where the parametrization returns admissible values of the physical $T$, we use the thermodynamic definition of temperature, namely, 
\bea
\frac{1}{T}=\frac{\partial S}{\partial U},
\eea
which can be computed by using Eqs.~(\ref{SandU}), after elimination of the $\beta^\star$ parameter. Figure~\ref{Fig:SandUab} shows the behavior of $S$ as a function of $U$ for $B/J=1$ and $q=0.2$, compared with the same functions when the thermal state $\hvrr$ is used and with the GB limit $q\to1$. In Fig.~\ref{Fig:SandUab}-(a) it is clear that the parametrization of Eq.~(\ref{Eq:RenormalizedBeta2}) leads to a entropy which is not a single-valued function of the internal energy. The black stars represent the positive values of temperature, which must be discarded (the same procedure needs to be performed with the branch of negative temperatures). The identification of those nonsense temperatures is displayed in Fig.~\ref{Fig:SandUab}-(b), implying that the physical temperature is the one when the parametrization becomes a monotonic increasing function of $\beta^\star$. The same situation is presented in Figs.~\ref{Fig:SandUab}-(c) and (d), for $B/J=4$ and $q=2$.

\begin{figure*}[h]
    \centering
    \includegraphics[scale=0.455]{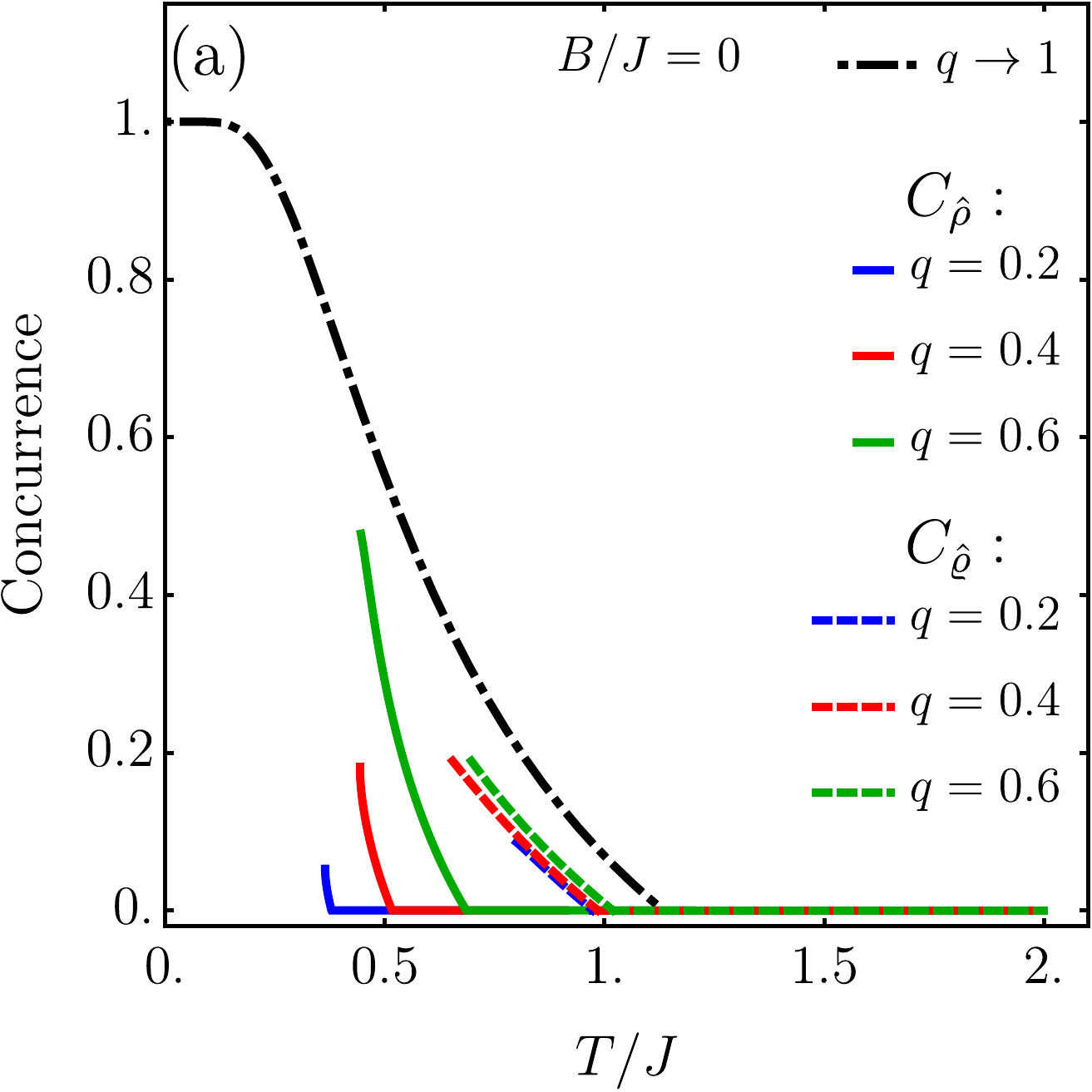}\hspace{0.25cm}
    \includegraphics[scale=0.455]{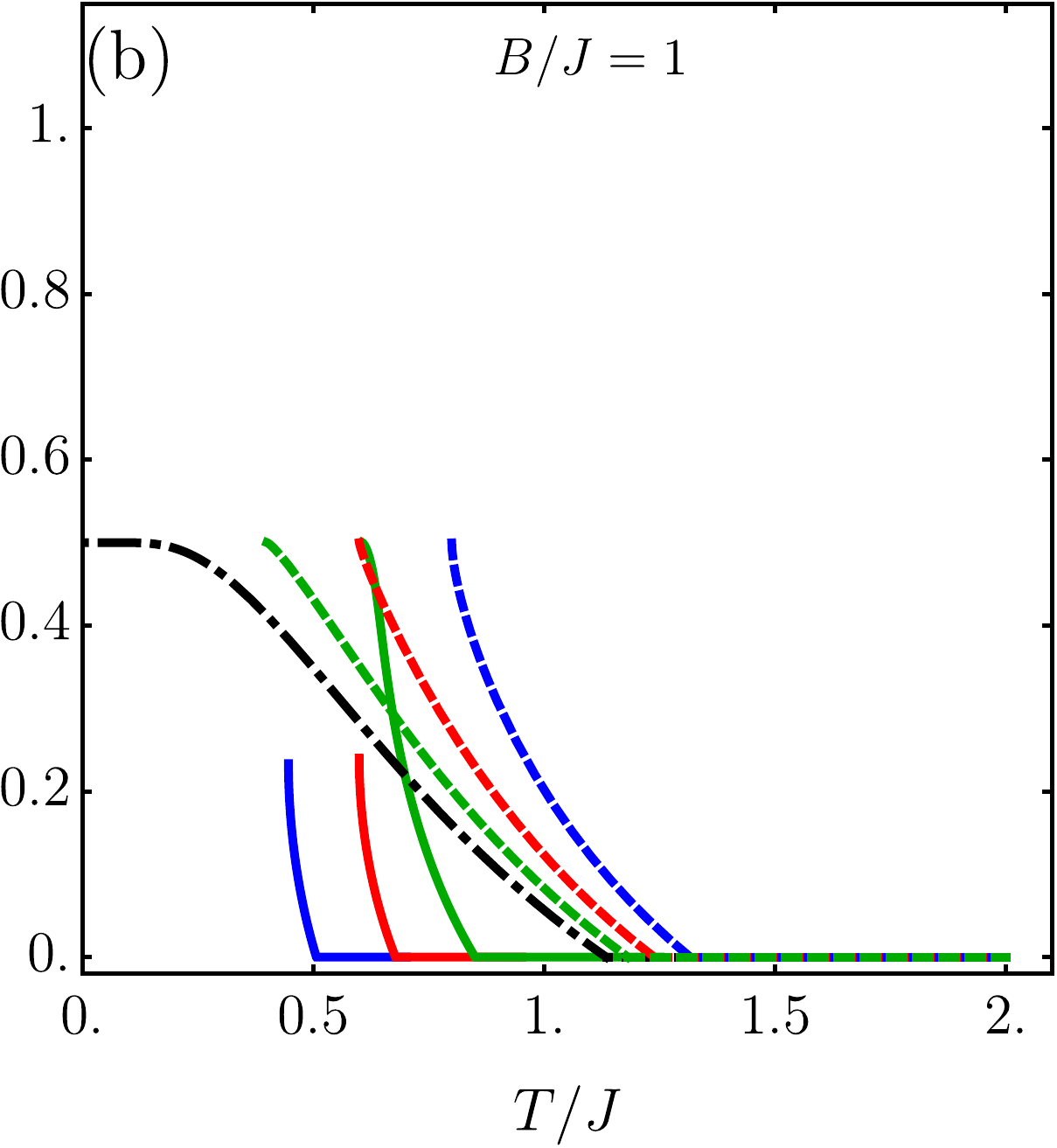}\hspace{0.25cm}
    \includegraphics[scale=0.455]{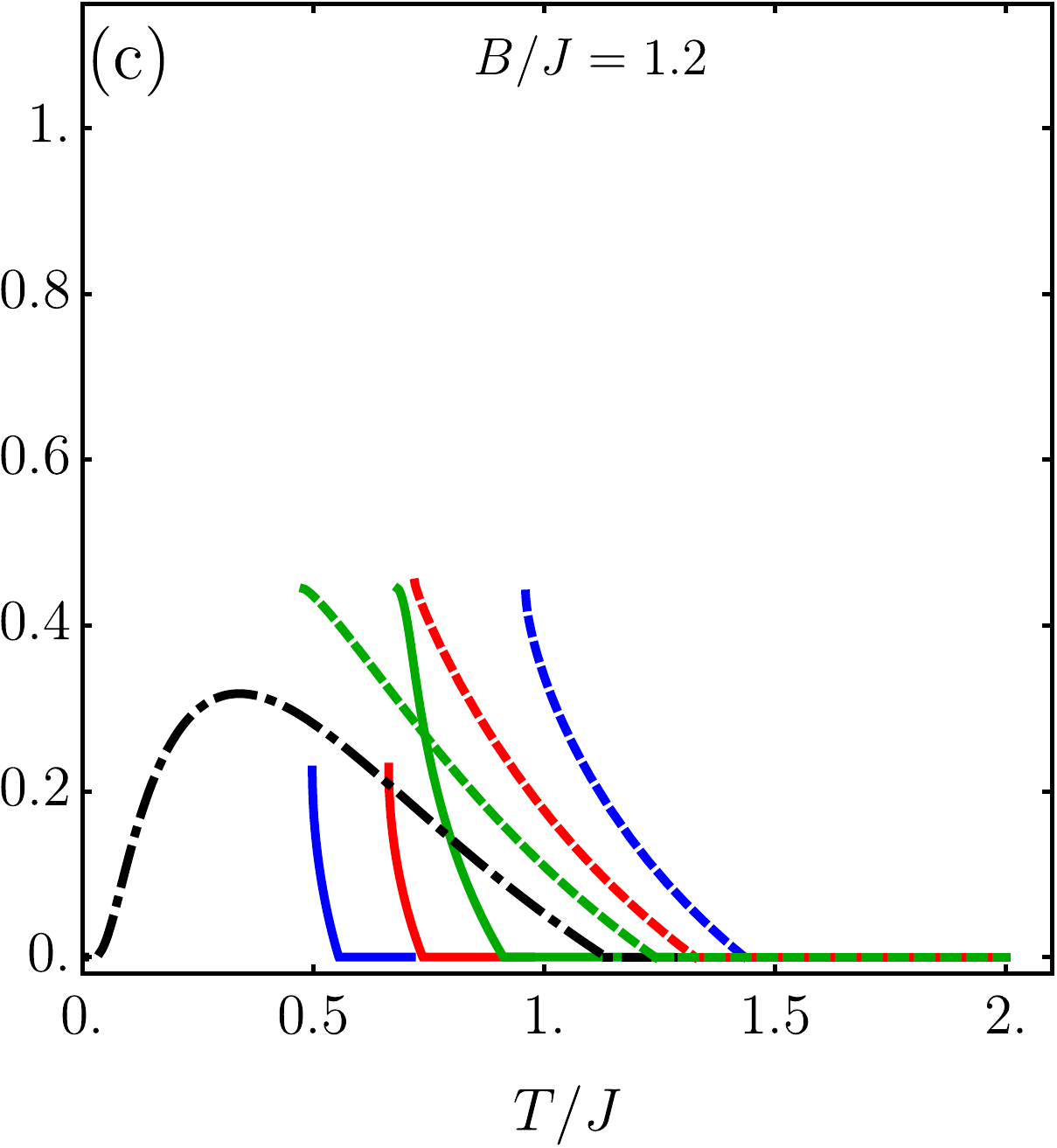}
    \caption{(Color online) Concurrence as a function of temperature computed from Eqs.~(\ref{Crho})-(\ref{CGB}) for $q<1$ and $B/J=0,1,1.2$.}
    \label{Fig:C1abc}
\end{figure*}
\begin{figure*}[h]
    \centering
    \includegraphics[scale=0.455]{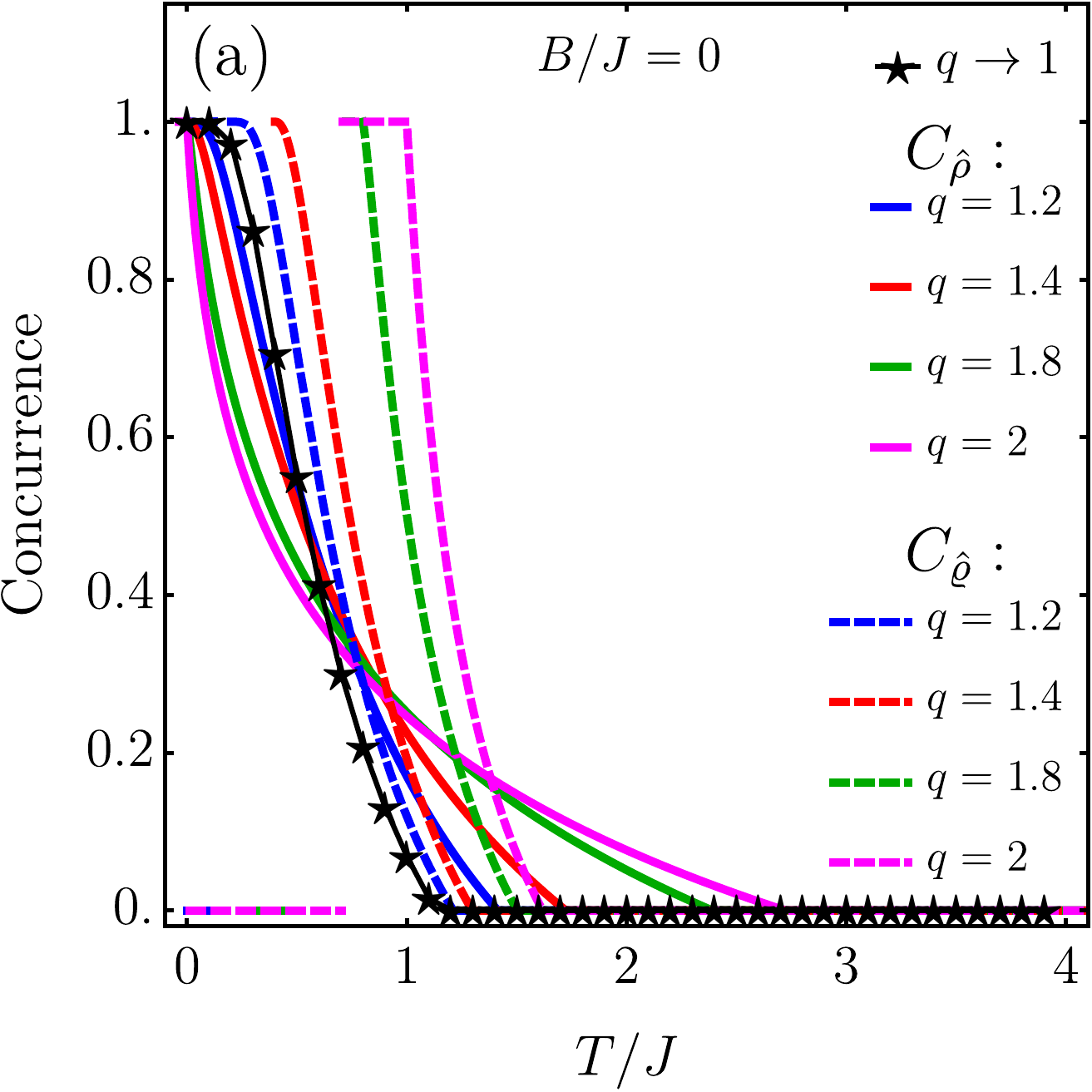}\hspace{0.25cm}
    \includegraphics[scale=0.455]{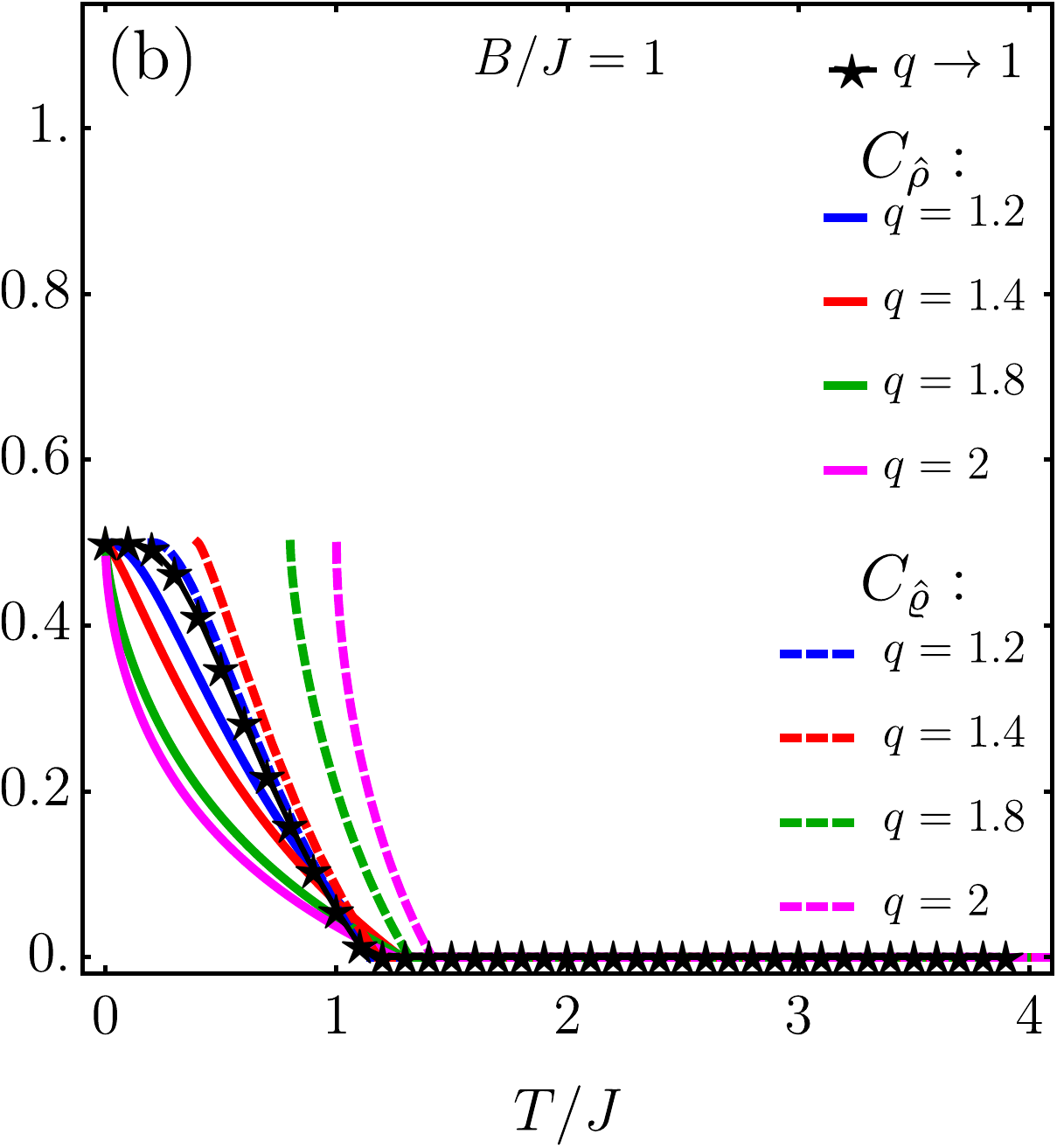}\hspace{0.25cm}
    \includegraphics[scale=0.455]{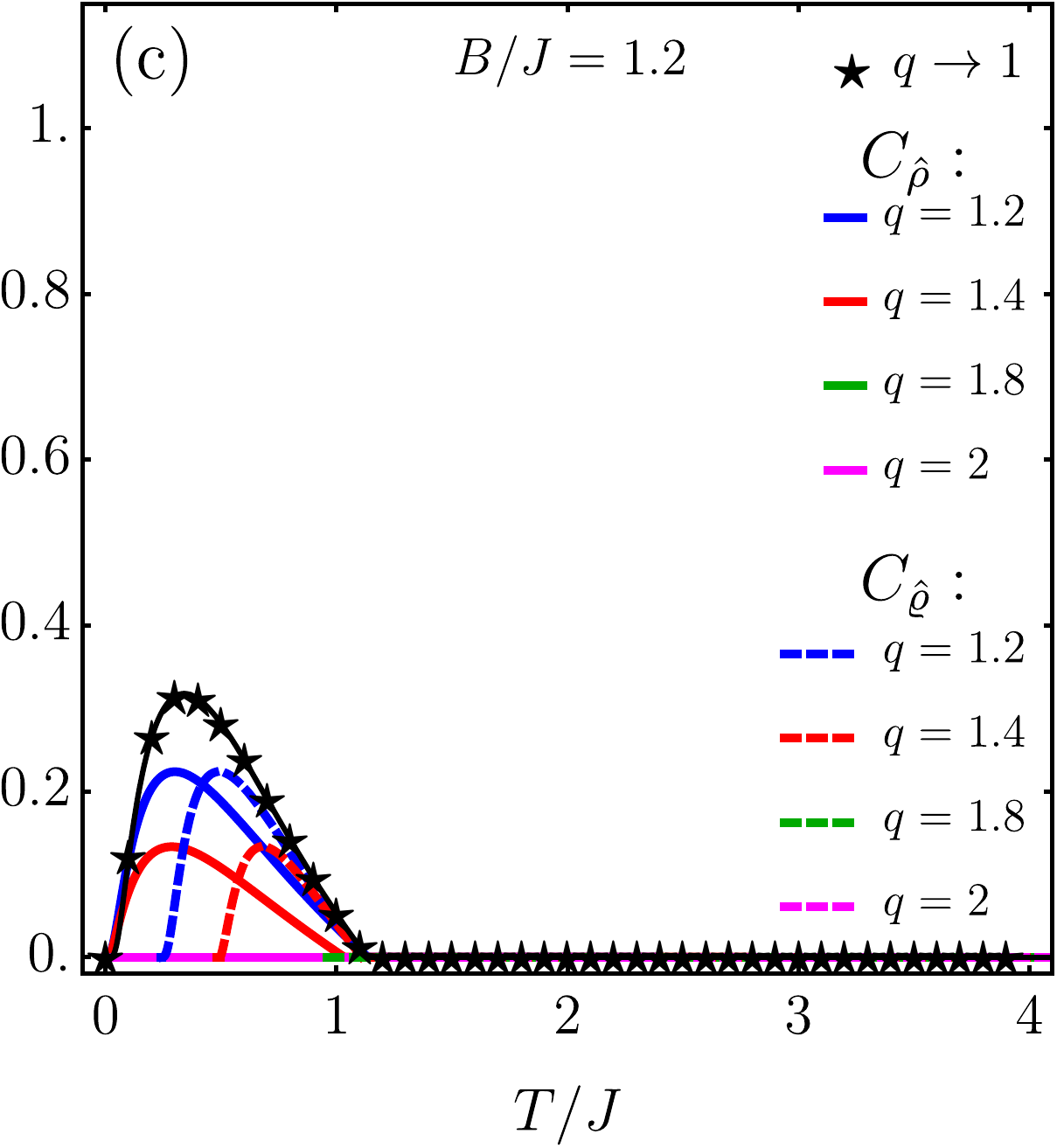}
    \caption{(Color online) Concurrence as a function of temperature computed from Eqs.~(\ref{Crho})-(\ref{CGB}) for $1<q\leq2$ and $B/J=0,1,1.2$.}
    \label{Fig:C2abc}
\end{figure*}
\begin{figure*}[h!]
    \centering
    \includegraphics[scale=0.455]{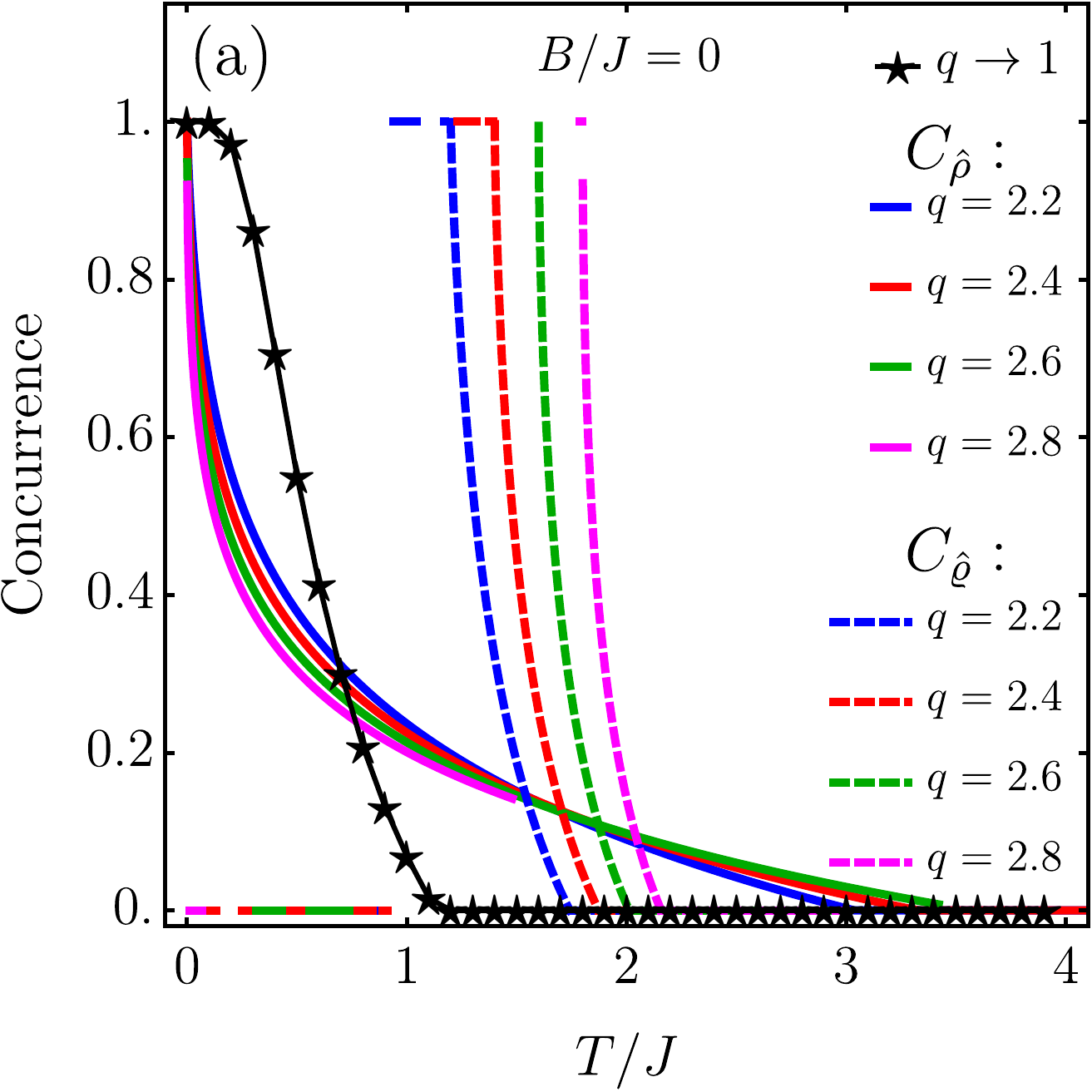}\hspace{0.25cm}
    \includegraphics[scale=0.455]{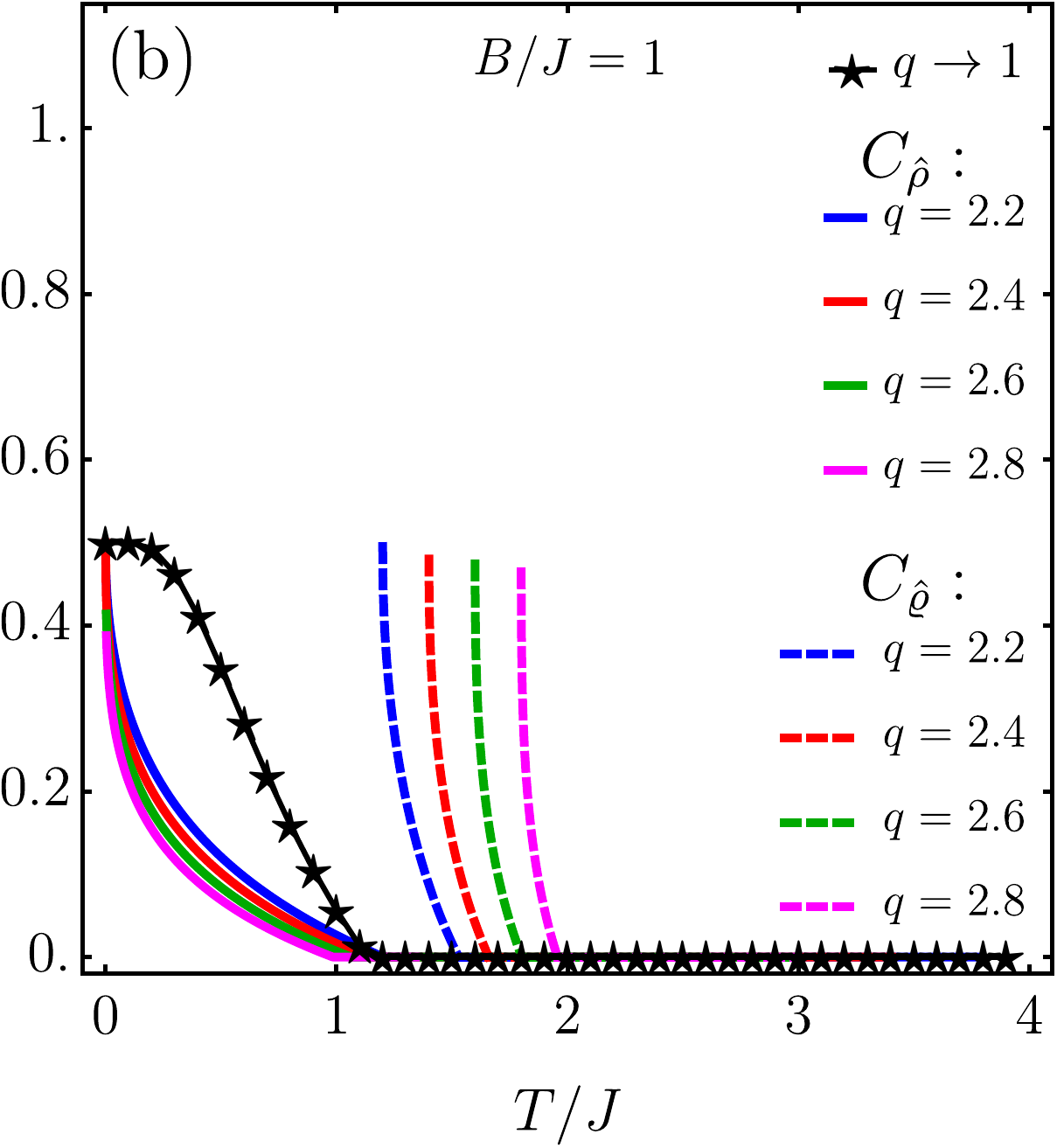}\hspace{0.25cm}
    \includegraphics[scale=0.455]{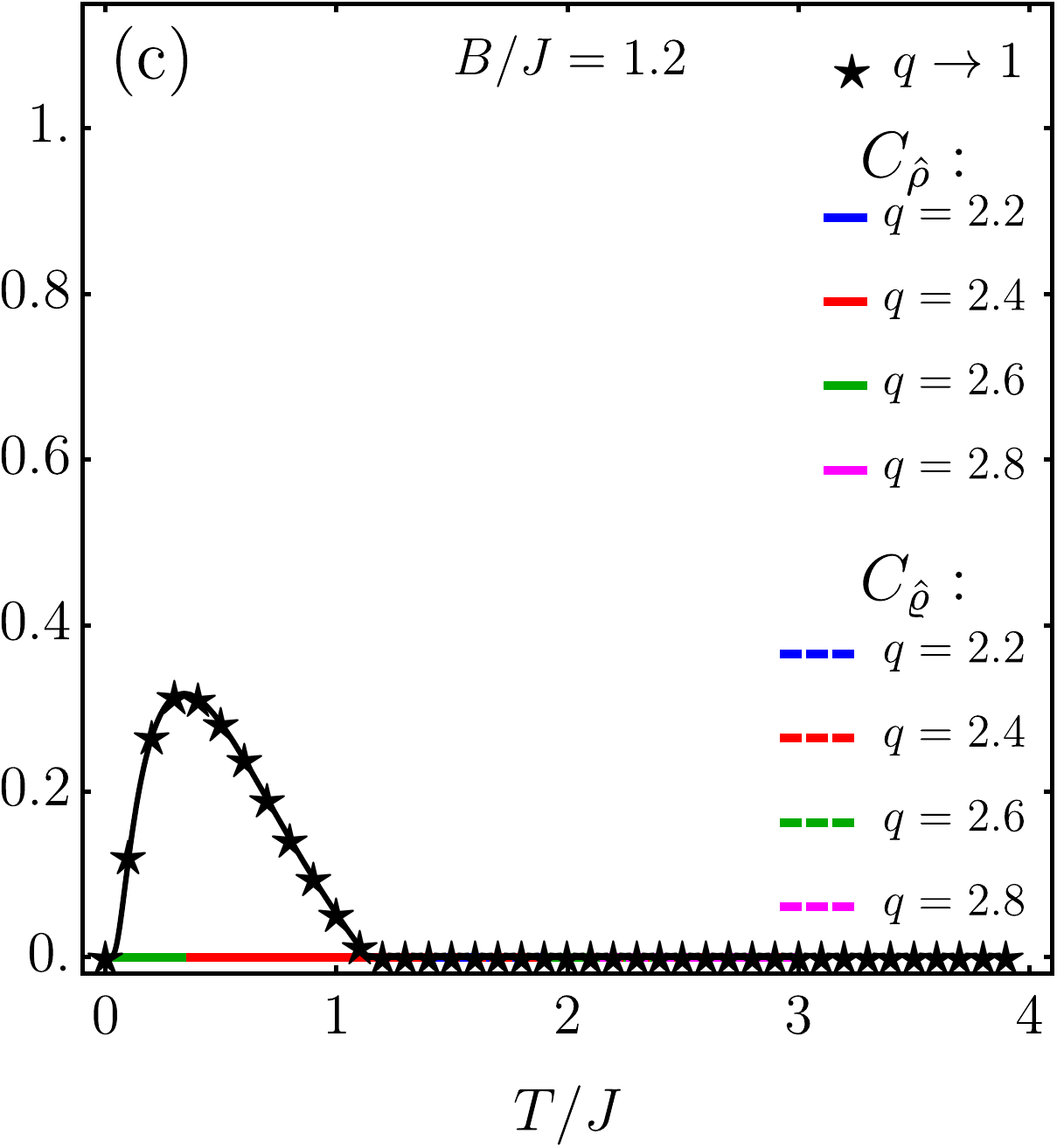}
    \caption{(Color online) Concurrence as a function of temperature computed from Eqs.~(\ref{Crho})-(\ref{CGB}) for $1\leq q\leq 2.8$ and $B/J=0,1,1.2$.}
    \label{Fig:C3abc}
\end{figure*}

\section{Concurrece}\label{Sec:Concurrence}

For a pair of qubits 1 and 2 with density matrix $\hrho_{12}$, the concurrence, as a measure of quantum entanglement, is defined as~\cite{PhysRevA.64.012313}:
\bea
C\equiv\max\left\{\lambda_1-\lambda_2-\lambda_3-\lambda_4,0\right\},
\label{ConcurrenceDef}
\eea
where the quantities $\lambda_1\geq\lambda_2\geq\lambda_3\geq\lambda_4$ are the squared root of the eigenvalues of the operator
\bea
\hat{\eta}_{12}=\hrho_{12}\left(\sigma_y\otimes\sigma_y\right)\hrho_{12}^*\left(\sigma_y\otimes\sigma_y\right),
\eea
so that if $C\neq0$ the qubits are entangled and $C=1$ corresponds to a maximally entangled state. We compare the concurrence when $\hrho_{12}$ is taken as $\hrho$, $\hvrr$ and the GB case $q\to1$, which are given by:
\bea
C_{\hrho}=\max\left\{\frac{\sinh_q[\beta^\star(\beta) J]-e_q[\beta^\star(\beta) B]e_q[-\beta^\star(\beta) B]}{\cosh_q[\beta^\star(\beta) J]+\cosh_q[\beta^\star(\beta) B]},0\right\},\nn\\
\label{Crho}
\eea
\bea
C_{\hvrr}=\max\left\{\frac{\sinh_q(\beta^\star J)-e_q(\beta^\star B)e_q(-\beta^\star B)}{\cosh_q(\beta^\star J)+\cosh_q(\beta^\star B)},0\right\},\nn\\
\label{Cvrho}
\eea
and
\bea
C_\text{GB}=\max\left\{\frac{\sinh(\beta J)-1}{\cosh(\beta J)+\cosh(\beta B)},0\right\}.
\label{CGB}
\eea

Note that we emphasize the dependence on the physical inverse temperature $\beta$ in Eq.~(\ref{Crho}) with the parametrization $\beta^\star(\beta)$. Also, our expression is different from Ref.~\cite{PhysRevE.95.042111}, given that in the $q$-calculus $e_q(x)e_1(-x)=e_q\left[(q-1)x^2\right]\neq1$.\\

Figure~\ref{Fig:C1abc} shows the concurrence as a function of temperature computed from Eqs.~(\ref{Crho})-(\ref{CGB}) for $q<1$ and different values of $B/J$. As can be noticed, the temperature renormalization shifts $C$ and gives a lower value compared with the case when the physical temperature definition is not taken into account (except for $B/J=0$). Furthermore, the critical temperature $T_c$ for $C_{\hrho}$, related to vanishing entanglement, moves away from the values obtained with $C_{\hvrr}$ as the magnetic field grows. Moreover, the concurrence enhances as $q$ gets closer to 1. Also,  Figs.~\ref{Fig:C1abc}-(b) and (c) show a hierarchy flip between the concurrence $C_{\hrho}$ for $q=0.2$ and $q=0.6$, compared with the behavior of $C_{\hvrr}$. \\


{ It is important to notice that we are not discussing the physical interpretation of the $q$-parameter. Then, it can take arbitrary values in the present work. Figures.~\ref{Fig:C2abc} and~\ref{Fig:C3abc} show the variations of the Concurrence functions $C_{\hat{\rho}}$ and $C_{\hat{\varrho}}$ when the temperature and the values of the $q$-parameter are modified. In contrast with Fig.~\ref{Fig:C1abc}(a), the Concurrence for $B/J=0$ in Figs.~\ref{Fig:C2abc}(a) and~\ref{Fig:C3abc}(a) are of the same order of the Concurrence when $q\to1$, reaching the maximum entangled state at $T=0$ for $C_{\hat{\rho}}$ and $T\neq 0$ for $C_{\hat{\varrho}}$. Moreover, the Concurrence curves for $C_{\hat{\rho}}$ are very close to each other, and the hierarchy (i.e., which curve is at the left or right of another) between them is inverse to the case computed from $C_{\hat{\varrho}}$. That inversion is due to the temperature renormalization provided by Eq.~(\ref{Eq:RenormalizedBeta2}. Finally, Figs.~\ref{Fig:C2abc}(c) and~\ref{Fig:C3abc}(c) show that $C_{\hat{\rho}}$ vanishes for higher values of $q$ and $B/J$. Indeed, for $q\geq3$, both $C_{\hat{\rho}}$ and $C_{\hat{\varrho}}$ are null, independent of the value of $B/J$. }


\section{Summary and conclusions}\label{Sec:summary}
In conclusion, in this work, we have studied the super-statistical description of { a spin-1/2 XX dimmer model} with an $\chi^2$ distribution for the temperature fluctuations, which allows an interpretation related to the non-additive Tsallis formalism. In order to preserve the Legendre structure of the thermodynamics, we followed the energy constraints that define the thermal state (or density operator) of the system with a physical temperature. Although the named constraints do not allow an analytical treatment, we used the fact that the density operator can be identified with a second set of energy constraints whose exact solution is well known through a temperature parametrization. Nevertheless, we found that such a parametrization leads to non-physical values of the temperature, and from analyzing the behavior of the entropy and internal energy, we imposed restrictions over the parametrization. 

In order to study the impact of the temperature fluctuations in the quantum entanglement by following a non-additive point of view, we computed the Concurrence as a measure of two qubits. We found that this quantity substantially changes with the thermal state prescription ($\hrho$ or $\hvrr$). Also, the critical temperature where the system is not entangled as well as the maximally entangled state is modified according to the super-statistical interpretation. The findings of this work can be implemented to understand in what situations the description provided by the super-statistical framework corresponds to the Tsallis formalism, and it can help to understand the role of the $q$-parameter better.  

\section*{Acknowledgements}
J. D. Casta\~no-Yepes acknowledges support from Consejo Nacional de Ciencia y Tecnolog\'ia CONCAyT (M\'exico) under grant number A1-S-7655. C. F. Ramirez-Gutierrez thanks academic group of TICA-UPQ (M\'exico) for modeling and simulation support. { The authors also thank Emiliano Adrián Rodríguez Reyes for a thorough reading of the manuscript and the language correction.}




\bibliography{bibQbitSE.bib}
\end{document}